# An empirical comparison of some outlier detection methods with longitudinal data

*Marcello D'Orazio*
*Italian National Institute of Statistics – Istat, Rome, Italy*

**Abstract**

*This note investigates the problem of detecting outliers in longitudinal data. It compares well-known methods used in official statistics with proposals from the fields of data mining and machine learning that are based on the distance between observations or binary partitioning trees. This is achieved by applying the methods to panel survey data related to different types of statistical units. Traditional methods are quite simple, enabling the direct identification of potential outliers, but they require specific assumptions. In contrast, recent methods provide only a score whose magnitude is directly related to the likelihood of an outlier being present. All the methods require the user to set a number of tuning parameters. However, the most recent methods are more flexible and sometimes more effective than traditional methods. In addition, these methods can be applied to multidimensional data.*

## 1. Introduction

Errors in data collected in sample surveys and administrative registers can affect the accuracy of the final estimates if they are not corrected. For this reason, National Statistical Institutes (NSIs) have always devoted significant resources to checking incoming data for errors. This process is known as *data editing* (also referred to as *statistical data editing* or input data validation; de Waal et al., 2011) and also aims to identify missing values, which are then replaced by *imputed values*[1]. The *imputation* process can also be used to replace erroneous values (deleted and imputed). A variety of statistical methods can be used in the editing and imputation sub-process, depending on many factors. These include the type of data source, the data collection mode, the nature of variables (continuous, categorical, or mixed-type) and the relationship existing between them, and the nature of errors and their potential impact on the final estimates.

This note focuses on the subset of data editing methods tailored to *outlier detection*. According to UNECE (2000)[2], "an outlier is a data value that lies in the tail of the statistical distribution of a set of data values", and the underlying idea is that "outliers in the distribution of uncorrected (raw) data are more likely to be incorrect". For example, when observing a single continuous variable such as household income, firm production or harvested area, an outlier may be the result of a measurement error; that is to say, the observed value is not the true value, which is not expected to be in the tail of the distribution. However, an outlier can also be an erroneous "extreme" value that, despite not being the result of a measurement error, may still have a significant impact on the final estimates and therefore require "special" treatment in the analysis[3].

---

[1] "Data editing and imputation" is a sub-process in the "Process" phase of the Generic Statistical Business Process Model GSBPM) https://statswiki.unece.org/display/GSBPM/GSBPM+v5.1
[2] "An outlier is an observation which is not fitted well by a model for the majority of the data. For instance, an outlier may lie in the tail of the statistical distribution or 'far away from the centre' of the data", Memobust Glossary, https://ec.europa.eu/eurostat/cros/content/glossary_en
[3] In sample surveys, in theory, it is possible to distinguish between a *representative outlier* (*i.e.* a unit in the sample that represents other units in the target population that have similar values) and *non-representative outlier* (when in the target population there are no other units showing similar values). https://ec.europa.eu/eurostat/cros/content/glossary_en

We consider traditional and recent approaches to outlier detection in the specific context of longitudinal data, where a continuous variable is observed on the same set of units at different times. This is the case for panel surveys in official statistics, which observe households, firms, or agricultural holdings. Section 2 briefly describes well-known outlier detection methods and illustrates some nonparametric approaches suggested in the fields of data mining and machine learning. These approaches have great potential when applied to this specific setting and, more generally, in official statistics. Section 3 compares the outcomes of applying the reviewed methods to panel data related to different types of statistical units, which are often investigated in surveys carried out by National Statistical Institutes (NSIs). Finally, Section 4 provides concluding remarks and suggests future areas of work.

## 2. Some outlier detection methods for longitudinal data

When a given quantitative non-negative variable *Y* is observed repeatedly over time on the same set of units, we can expect a high correlation between subsequent measurements; this feature represents a useful information in setting up an efficient outlier detection procedure. It becomes even more relevant when the objective is the estimation of the change over time of a population parameter related to *Y*.

Formally, let $(y_{t1}, y_{t2}, \ldots, y_{tn})$ be the values of *Y* observed at time *t* on a set of *n* units, being $y_{ti} \geq 0$; the ratio $r_i = y_{t_2 i}/y_{t_1 i}$ denotes the "individual change" from time $t_1$ to time $t_2$ for unit *i*, being $t_2 > t_1$. Data editing literature suggests various methods to check whether the individual change ($r_i$) is too large or too low (*see e.g.* the theme "Editing for Longitudinal Data" in Eurostat, 2014); a very popular one was suggested by Hidiroglou and Berthelot (1986) and its characteristics are summarised in Sub-section 2.1

### 2.1 Hidiroglou-Berthelot method for outlier detection

Hidiroglou and Berthelot (1986) suggest to look at the empirical distribution of the ratios $r_i = y_{t_2 i}/y_{t_1 i}$ ($i = 1, 2, \ldots, m$; being $m \leq n$, after discarding 0s and missing values, if any, in both $y_{t_2 i}$ and $y_{t_1 2}$). In particular, they first transform the ratios in the following manner:

$$s_i = \begin{cases} 1 - \dfrac{r_M}{r_i}, & \text{if} \quad 0 < r_i < r_M \\[6pt] \dfrac{r_i}{r_M} - 1, & \text{if} \quad r_i \geq r_M \end{cases} \qquad (1)$$

such that $s_i = 0$ when a ratio is equal to the median of ratios ($r_i = r_M$); then, to account for the magnitude of data and give more "importance" to units involving high values of the *Y* variable, suggest to derive the following *score*:

$$E_i = s_i \bigl[\max(y_{t_1 i}, y_{t_2 i})\bigr]^U \qquad (2)$$

where *U* can range from 0 to 1 ($0 \leq U \leq 1$) and controls the role of the magnitude in determining the importance associated to the centered ratios; a common choice consists in setting $U = 0.5$.

Identification of potential outliers relies on the assumption that the scores are approximately distributed according to a Gaussian distribution; in practice, the parameters of Gaussian distribution are estimated using robust methods and units outside the interval:

$$[E_M - C \times d_{Q1}, E_M + C \times d_{Q3}] \qquad (3)$$

are identified as potential outliers. In expression (3):

$$d_{Q1} = max(E_M - E_{Q1}, |A \times E_M|) \quad d_{Q3} = max(E_{Q3} - E_M, |A \times E_M|) \qquad (4)$$

being $E_{Q1}$, $E_M$ and $E_{Q3}$ the quartiles of the *E* scores. The constant $A$ is a positive small quantity (suggested $A = 0.05$) introduced to overcome cases of $E_M = E_{Q3}$ or $E_M = E_{Q1}$ that may occur when the ratios are too concentrated around their median. The parameter *C* determines how far from the median the bounds should be; commonly suggested values are $C = 4$ or $C = 7$ but larger values can be considered, depending on the tails of the distribution of the *E* scores. In practice, the bounds (3) allow for a slight skewness in the distribution of the *E* scores.

Recently, Hidiroglou and Emond (2018) suggested to replace $E_{Q1}$ and $E_{Q3}$ with respectively the percentiles $E_{P10}$ and $E_{P90}$ when a large proportion of units (>1/4) share the same value of the ratio, since in this case the "standard" method would detect too many observations as potential outliers.

Practically, the decision about the values of the "tuning" constants *U* and *C* is not straightforward and requires a graphical investigation of the distribution of the scores and different attempts with alternative values of both the constants. A useful practical suggestion is to start inspecting the (suspicious) ratios by sorting them in decreasing order with respect to the absolute value of the score ($|E_i|$). Hidiroglou and Emond (2018) suggest also an additional graphical inspection procedure.

In the R environment (R Core Team 2022) the Hidiroglou Berthelot (HB) procedure is implemented by the function `HBmethod()` available in the package **univOutl** (D'Orazio 2022) that includes also graphical facilities for inspecting the scores, in line with Hidiroglou and Emond (2018) recommendation. In addition, **univOutl** has facilities to identify outliers in univariate case with methods based on robust location and scale estimates of the parameters of the Gaussian distribution.

## 2.2 Nonparametric methods

The nonparametric methods for outlier detection are very popular because they do not introduce an explicit assumption on the underlying distribution. For sake of simplicity, in describing these methods it is considered the problem of detecting outliers when a generic continuous variable *X* is observed on set of *m* observations. The following Sub-sections describe some well-known approaches using boxplots as well as recent proposals in the field of data mining and machine learning and, in particular, *distance* and *tree-based methods*.

### 2.2.1 Outliers detection with boxplots

Drawing a *boxplot* (*box-and-whisker* plot) is a popular approach to outlier detection:

$$[Q1_x - c \times IQR_x; \ Q3_x + c \times IQR_x] \qquad (5)$$

where $IQR_x$ is the inter-quartile range ($IQR_x = Q3_x - Q1_x$) and, usually, $c = 1.5$, 2 or 3; units outside the bounds (*whiskers*) are considered outliers.

To account for moderate skewness Hubert and Vandervieren (2008) suggested an "adjusted" boxplot:

$$[Q1_x - 1.5 \times exp(aM) \times IQR_x; \ Q3_x + 1.5 \times exp(bM) \times IQR_x] \qquad (6)$$

being *M* the *medcouple* measure of skewness ($-1 \leq M \leq 1$; Brys *et al.* 2004) that when greater than 0 indicates positive skewness and requires setting $a = -4$ and $b = 3$ in expression (6); on the contrary, with negative skewness ($M < 0$) it is suggested to set $a = -3$ and $b = 4$. The authors claim that the adjusted boxplot fences in (6) work with moderate skewness, *i.e.* $-0.6 \leq M \leq 0.6$. Unfortunately, the adjusted boxplot permits only to set $c = 1.5$ and it is not possible to use alternative values.

D'Orazio (2022) implemented standard and adjusted boxplot in the function `boxB()` of the R package **univOutl**. In addition, the function `HBmethod()` permits to apply the adjusted boxplot to the *E* scores.

### 2.2.2 Distance-based outlier detection

The idea of using distance measures in outlier detection is a direct consequence of the fact that we search for observations that are far from the centre of the data. In practice, distance-based outlier detection methods search for an observation that has very few other observations close to it, the less are the observations close to a unit the higher will be the chance that it is an outlier. Knorr and Ng (1998) suggest to identify as outlier the observation that has less than *k* observations at a distance less than or equal to a threshold $\delta$. This approach requires deciding: (i) how to measure the distance, (ii) the distance threshold $\delta$ and, (iii) the *k* parameter. The first two choices are strictly related and relatively simple in the univariate setting (but not in the multidimensional case) as different distance functions may leave unchanged the set of nearest neighbors of a given unit.

Knorr and Ng (1998) approach does not provide a ranking for the potential outliers. To overcome this difficulty Ramaswamy *et al.* (2000) suggest identifying the potential outliers by calculating the *k* nearest neighbor (*k*-NN) distance; in practice, if $d_i^{(k)}$ is the distance of the *i*th from its *k*-nearest neighbor, the units showing largest values of $d_i^{(k)}$ are potential outliers. This simple approach can be computationally demanding in the presence of many observations and variables, anyway there are algorithms that simplify the search (see *e.g.* Hautamäki *et al.* 2004); the problem becomes much simpler in the univariate case where the initial ordering of the units reduces the computational effort.

The *k*-NN distance proposed by Ramaswamy *et al.* (2000) is a very popular approach and many variants are available. A well-known extension assigns to each unit a "weight" consisting in the sum of its distance from the corresponding *k* nearest neighbor observations (Angiulli and Pizzuti 2002):

$$\omega_i^{(k)} = \sum_{j=1}^{k} d_i^{(j)} \qquad (7)$$

(Hautamäki *et al* 2004 suggest using the average). Common choices for the parameter *k* are 5 or 10, however literature does not indicate a rule-of-thumb. Campos *et al.* (2016) note that the sum of distances makes the scores less sensitive to the value of *k*. Obviously, if *k* is too large, then the weight may become quite large even for non-outlying observations, since, as shown by expression (7), the final score will be the result of a sum of higher number of terms.

In general, distance-based outlier detection methods are strictly related to outliers' detection based on kernel density estimation techniques; this note will not address such methods, but it is worth noting that when *k*-NN is applied to density estimation problems a possible rule-of-thumb consist in setting $k = ceiling(\sqrt{m})$ and, more in general $k \sim m^{4/5}$.

The main drawback with *k*-NN methods is that they do not identify directly the potential outliers, like boxplots or HB method, rather they provide a summary score (distance or "weight") whose magnitude indicates the chance of being an outlier; the largest is the score the higher is the chance that a given observation is an outlier. To identify a possible threshold such that observations with a score above the threshold are identified as potential outliers, Hautamäki *et al.* (2004) suggest:

$$u_0 = \varepsilon \times max\left[u_{(i+1)}^{(k)} - u_{(i)}^{(k)}\right], \quad i = 1, 2, \ldots, m-1 \quad (8)$$

where $u_i^{(k)} = \omega_i^{(k)}$ or $u_i^{(k)} = d_i^{(k)}$, $u_{(i+1)}^{(k)} \geq u_{(i)}^{(k)}$ and $\varepsilon$ is a user-defined constant $0 < \varepsilon < 1$. This rule, introduces an additional parameter to set ($\varepsilon$); practically, a graphical inspection of the ordered scores $u_i^{(k)}$ can be more effective: once sorted them increasingly, good candidate thresholds the values corresponding to "jumps" in the plot (abnormal increase in the score).

The Knorr and Ng (1998) approach is closely related to the DBSCAN (*Density-based spatial clustering applications with noise*; Ester *et al* 1996) clustering algorithm where the observations not "reachable" by any other observation are identified as *noisy* observations (outliers). The "reachability" depends on a distance threshold $\delta$; in practice two observations *i* and *j* are *directly reachable* if their distance is less or equal than $\delta$ ($d_{i,j} \leq \delta$), while they are only *reachable* if there is a path of three of more observations to go from *i* to *j*, where each couple of units in the path is directly reachable. The DBSCAN algorithm requires setting also a value for *g* that is needed to identify the *core observations*, *i.e.* observations that have at least $k = g - 1$ distinct units at a distance smaller or equal to $\delta$. To identify a value for $\delta$ it is suggested to plot the *k*-NN distances in increasing order and set $\delta$ equal to the distance where the plot shows a jump.

In R some distance-based methods for outlier detection are implemented in the package **DDoutlier** (Madsen 2018), although *k*-NN distance is calculated in many other R packages; the package **dbscan** (Hahsler *et al*. 2019 and 2022) implements the DBSCAN clustering algorithm but has also facilities to efficiently calculate the *k*-NN distance.

*2.2.3 Outlier detection with isolation forest*

*Isolation forest* is an unsupervised decision-tree-based algorithm that consists in fitting an ensemble of *isolation trees* (Liu *et al.* 2008 and 2012). The underlying idea is that outlying observations have a higher chance of being separated by the other ones in one branch of the partitioning tree with relatively few splits. In the univariate case an arbitrary threshold $x_o$ is selected at random within the range of $X$ ($[min(x_i), max(x_i)]$) and all the observations are divided into two groups according to whether they show higher or lower values than $x_o$. This randomised splitting process is applied recursively (*i.e.* divide the units into two groups then repeat the process in each group, and so on) until no further split is possible or until meeting some other criteria. The final outcome is an isolation tree where the more observations show similar *X* values, the longer (more splits) it will take to separate them in small groups (or alone) compared to less occurring *X* values; for this reason, the *isolation depth* (number of splits needed to isolate a unit) can be considered as a tool for detecting outliers.

Since the isolation depth estimated in a single isolation tree would be characterised by a high variability, its reduction can be achieved by building an ensemble of isolation trees – the isolation forest – and then derive the final score by averaging over the fitted trees. As in random forests, each single isolation tree is fit on a bootstrap sample of *q* ($q < m$) observations randomly selected. In the Liu *et al* proposal (2008 and 2012) the partitioning stops when a node has only one observation or all units in a node have the same values (in some cases it is also introduced a maximum value for the tree height, *e.g.* $l_{max} = ceiling(log_2(q))$).

Formally, if $h(x_i)$ is the *path-length* or *depths*, *i.e.* the number of splits to reach the *i*-th observation in a fitted tree, Liu *et al.* (2008 and 2012) suggest to associate to each observation the following score:

$$u_i = 2^{-\frac{E[h(x_i)]}{c(q)}} \qquad (9)$$

Where $E[h(x_i)]$ is the average path length across the ensemble of the fitted trees and

$$c(q) = 2 \times H(q-1) - 2\frac{q-1}{q} \qquad (10)$$

being $H(\cdot)$ the *harmonic number*. The Authors demonstrate that the resulting score ranges from 0 to 1 ($0 < u_i \leq 1$) being a monotonic function of $h(x_i)$ and, in particular

– when $E[h(x_i)] \to q-1$ then $u_i \to 0$;
– when $E[h(x_i)] \to c(q)$ then $u_i \to 0.5$;
– when $E[h(x_i)] \to 0$ then $u_i \to 1$.

Practically, scores close to 1 indicate observations with a very short average path length that tend to be isolated earlier than the other ones and therefore denote outlying observations. As a consequence, setting a threshold score $u_0$ will return as outliers all the units having a score $u_i > u_0$. Generally, it is suggested to consider $u_0 > 0.5$ but a graphical inspection of the ordered scores can be beneficial in deciding $u_0$.

The isolation forest is very efficient and can handle multi-modal distributions. It requires setting two tuning parameters, the subsample size $q$ and the number of trees to fit. In the first case the Liu *et al* (2008 and 2012) claim that even a small subsample size ($q = 256$) can work with very large data-sets, while at least 100 trees should be fitted; this latter number should be increased when the achieved scores are on average quite below 0.5, as this may point out a problem of unreliable estimation of the average path length. It is worth noting that the standard method proposed by Liu *et al.* (2008 and 2012) is developed also to handle outlier detection in a multidimensional framework, where the creation of each tree requires a recursive random selection of one of the available variables and of the corresponding random splitting point. In the univariate case, with a single variable there is just the random selection of the splitting point and consequently there is no need to grow a high number of trees. It is worth noting that, in the multivariate case the standard algorithm (Liu *et al* 2008 and 2012) practically would consist in an ensemble of results related to the application of isolation forest independently variable by variable. To compensate this drawback, Hariri et al (2018) proposed an *extended isolation forest* that in the branching step considers jointly two or more variables; for instance, when two variables are randomly selected then the algorithm partitions repeatedly the units according to a regression line whose intercept and slope are randomly generated each time.

In R the standard isolation forest is implemented in the package **solitude** (Srikanth 2021) while the package **isotree** (Cortes 2022) implements the "base" isolation forest algorithm as well as some of its variants.

## 3. Application of the chosen methods to some data from panel surveys

This section investigates the performances of the methods presented in the previous Section when applied to different datasets related to panel or pseudo-panel surveys that are described in Table 1. In practice, in each dataset the HB procedure is applied to the chosen variable and the resulting $E_i$ scores, calculated using expression (2), become the input of the outlier detection techniques listed in the first column of Table 2, whose corresponding parameters/tuning constants are given in the second column of the table. All the analyses were carried out in the R environment, the columns 3 and 4 in Table 2 provide details related to the chosen R package, function and corresponding arguments (arguments not explicitly mentioned are set equal to their default values)[4].

**Table 1 – Datasets used in the experiments**

| Dataset/survey | Number of units | Type of units | Description |
|---|---|---|---|
| RDPerfComp | 509 | Firms | R&D performing US manufacturing; yearly observations from 1982 to 1989 of the following variables: production, labour and capital[5] |
| RiceFarms | 171 | Farms | Indonesian rice farm dataset, 171 farms producing rice observed 6 times. Many variables are available: hectares of cultivated area, gross output of rice in kg, net output, etc.[6] |
| Wages | 595 | Individuals | A panel of 595 individuals from 1976 to 1982, taken from the Panel Study of Income Dynamics (PSID); many variables available (see footnote 2) |
| Survey on Household Income and Wealth (SHIW) | 3804 | Households | Subset of panel households observed in years 2014 and 2016. Many variables available: net income, consumptions, wealth, etc.[7] |

**Table 2 – Methods, R functions and corresponding tuning parameters**

| Method | Parameters | R function and package | Arguments in the R function |
|---|---|---|---|
| Hidiroglou-Berthelot ("HB") | Quartiles (deciles for Wages dataset) $U = 0.5$; $C = 7$; $A = 0.05$ | `HBmethod()` in **univOutl** | `U=0.5, A=0.05, C=7, pct=0.25` (`pct=0.10` for Wages dataset) |
| Skewness-adjusted boxplot ("SABP") (see eqn. (6)) | | `boxB()` in **univOutl** | `k=1.5, method='adjbox'` |
| Isolation Forest ("IF") | No subsampling; 500 trees | `isolation.forest()` in **isotree** | `ntrees=500` |
| DBSCAN | Three runs with different values of $g$ (6, 11,16) and different thresholds for $\delta$ (decided after graphical inspection of the sorted $(g\text{-}1)$-NN distances calculated on the $E_i$) | `dbscan()` in **dbscan** | `minPts=6, minPts=11, minPts=16` `eps` set equal to the decided $\delta$ for each combination of `minPts` and the various datasets |
| $k$-NN outlier detection ("$k$-NN-dist") | Three runs with different values of $k$ (5, 10,15) | `kNNdist()` in **dbscan** | `k=5, k=10, k=15` |
| $k$-NN weights ("$k$-NN-weight"; see expression (7)) | Three runs with different values of $k$ (5, 10,15) | `kNNdist()` in **dbscan** | `k=5, k=10, k=15, all=TRUE` |

The variable inspected in RDPerfComp dataset is the firms' production of year 1983 vs. year 1982. Figure 1 reports the observed scatterplot (1a); plot (1b) shows the distribution of the HB scores ($E_i$) and the vertical continuous lines indicate the HB bounds provided by (3) while the dashed lines are the fences of the skewness-adjusted boxplot, provided by (6). The histogram shows a moderate negative skewness ($M = -0.2338$) and the SABP fences identify a higher number of potential outliers if compared to HB (with $C = 7$ and $A = 0.5$).

The right-side plot (1c) reports the scatterplot of the IF scores ($u_i$) vs. $E_i$. Many observations identified as outliers by HB method have an IF score slightly above 0.5, while the observations with the maximum observed IF score (slightly beyond 0.8, being 1 the maximum achievable score) are quite few.

---

[4] The used R code can be found in the GitHub repository: *link to be included*
[5] https://www.nuffield.ox.ac.uk/users/bond/index.html. See also the R package **pder** https://CRAN.R-project.org/package=pder
[6] R package **plm** https://cran.r-project.org/package=plm

Bank of Italy, Survey on Household Income and Wealth[7], years 2014 and 2016. Public use anonymised microdata distributed for research purposes https://www.bancaditalia.it/statistiche/tematiche/indagini-famiglie-imprese/bilanci-famiglie/distribuzione-microdati/index.html?com.dotmarketing.htmlpage.language=1

**Figure 1 – Scatterplot of the data related to firms' production (1a), distribution of the HB scores (1b) and relation between HB and IF scores (1c)**

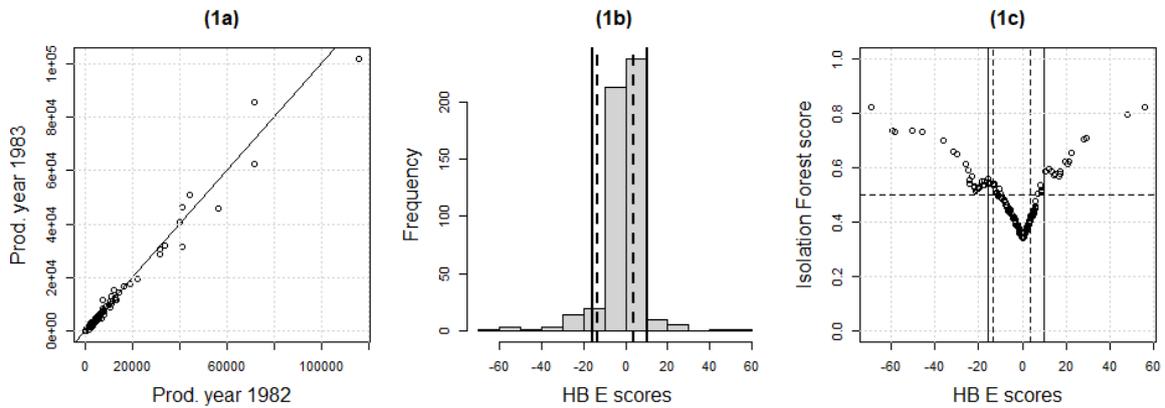

**Figure 2 – Firms' production data, relationship between HB and scores provided by the *k*-NN methods**

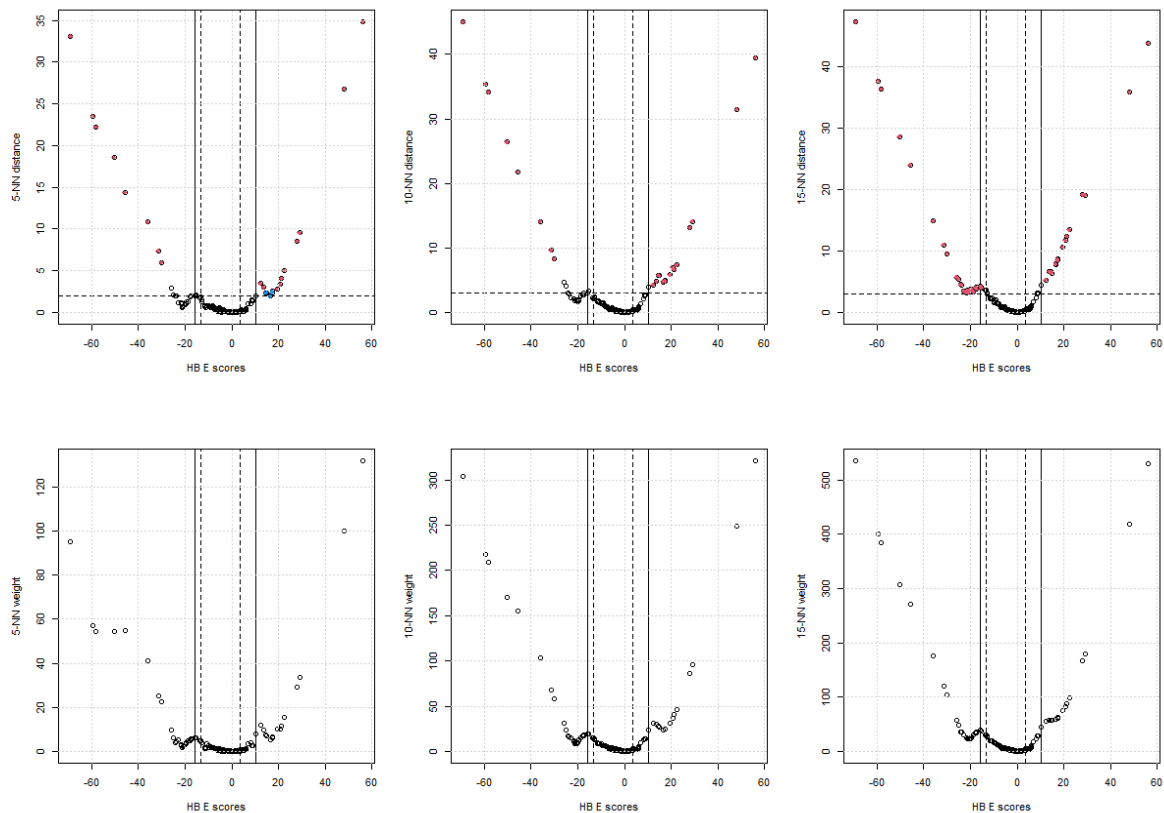

Figure 2 reports the scatterplots of scores provided by *k*-NN-dist (2a-2c) and *k*-NN-weight (2d-2e) compared to the input HB scores ($E_i$). Scatterplots (2a-2c) show also the findings of DBSCAN with respectively $\delta = 2$, $\delta = 3$ and $\delta = 3$; in these plots the red-color points indicate the noisy points (outliers), while the blue-color ones form a separate cluster of observations, far for most of the observations that however are not identified as outliers. In this example, the outliers returned by DBSCAN are always less than those provided by the standard HB method. More in general 5-NN and 10-NN help more than 15-NN distance in identifying potential outliers (units with higher distance).

Scatterplot (2d), (2e) and (2f) compare the $E_i$ with the sum of the *k*-NN distances (*k*-NN-weight). If compared to the "standard" *k*-NN-dist, the sum of the distances (weight) as expected seems less sensitive to the value of *k* (cf. Campos *et al.* 2016) and helps more in detecting the potential outliers (units with the highest weight $\omega_i^{(k)}$), in particular when $k = 5$ and $k = 10$.

Table 3 shows the estimated Kendall's *tau* correlation coefficient between the various scores obtained at the end of the different procedures for outlier detection (for HB it is considered the absolute value $|E_i|$). Correlations are quite high indicating a high concordance between rankings of the scores produced by the different methods. IF scores are highly correlated with $|E_i|$. Concordance between $|E_i|$ and the scores provided by the application of *k*-NN methods increases with increasing values of *k*.

**Table 3 – Kendall's correlation between the scores assigned to the firms**

|  | IF | 5-NN-dist | 10-NN-dist | 15-NN-dist | 5-NN-weight | 10-NN-weight | 15-NN-weight |
|---|---|---|---|---|---|---|---|
| \|E\| | 0.8984 | 0.7672 | 0.8269 | 0.8689 | 0.7435 | 0.8094 | 0.8471 |
| IF |  | 0.8091 | 0.8665 | 0.9009 | 0.7845 | 0.8544 | 0.8920 |
| 5-NN-dist |  |  | 0.829 | 0.7974 | 0.9045 | 0.8931 | 0.8587 |
| 10-NN-dist |  |  |  | 0.8913 | 0.8050 | 0.9198 | 0.9414 |
| 15-NN-dist |  |  |  |  | 0.7717 | 0.8596 | 0.9144 |
| 5-NN-weight |  |  |  |  |  | 0.8712 | 0.8296 |
| 10-NN-weight |  |  |  |  |  |  | 0.9400 |

Figures 3 and 4 summarize the results of the different outlier detection procedures when applied to the area harvested to rice production in the observed farms (5[th] time occasion vs. the 4[th]) listed in the RiceFarm dataset. Histogram (3b) indicates a moderate positive skewness ($M = 0.3697$). The fences of the SABP are close to the bounds of HB intervals, in particular on the right tail of the distribution; as expected, SABP seems to better account for moderate positive skewness. The identified outlying farms are relatively few and, in general, show an IF score greater than 0.6 (with few exceptions, located in the left tail). DBSCAN with the chosen distance thresholds ($\delta = 1$, $\delta = 1.5$ and $\delta = 2$, respectively; decided by after graphical inspection of the sorted *k*-NN distances) identifies quite few outliers, slight less than those identified by HB or SABP. Plots related to *k*-NN methods (4a-4c and 4d-4f) show that there are few outlying farms with a score (*k*-NN distance or *k*-NN weight) not close to that of the majority of farms.

**Figure 3 – Scatterplot of the area for rice production (3a), distribution of the HB scores (3b) and relation between HB and IF scores (3c)**

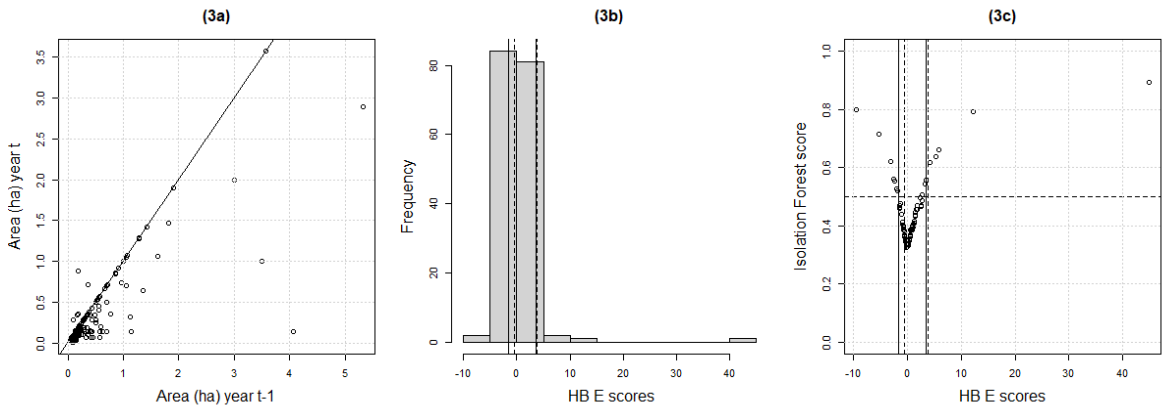

**Figure 4 – Rice-growing area in farms, relationship between HB and scores provided by the *k*-NN methods**

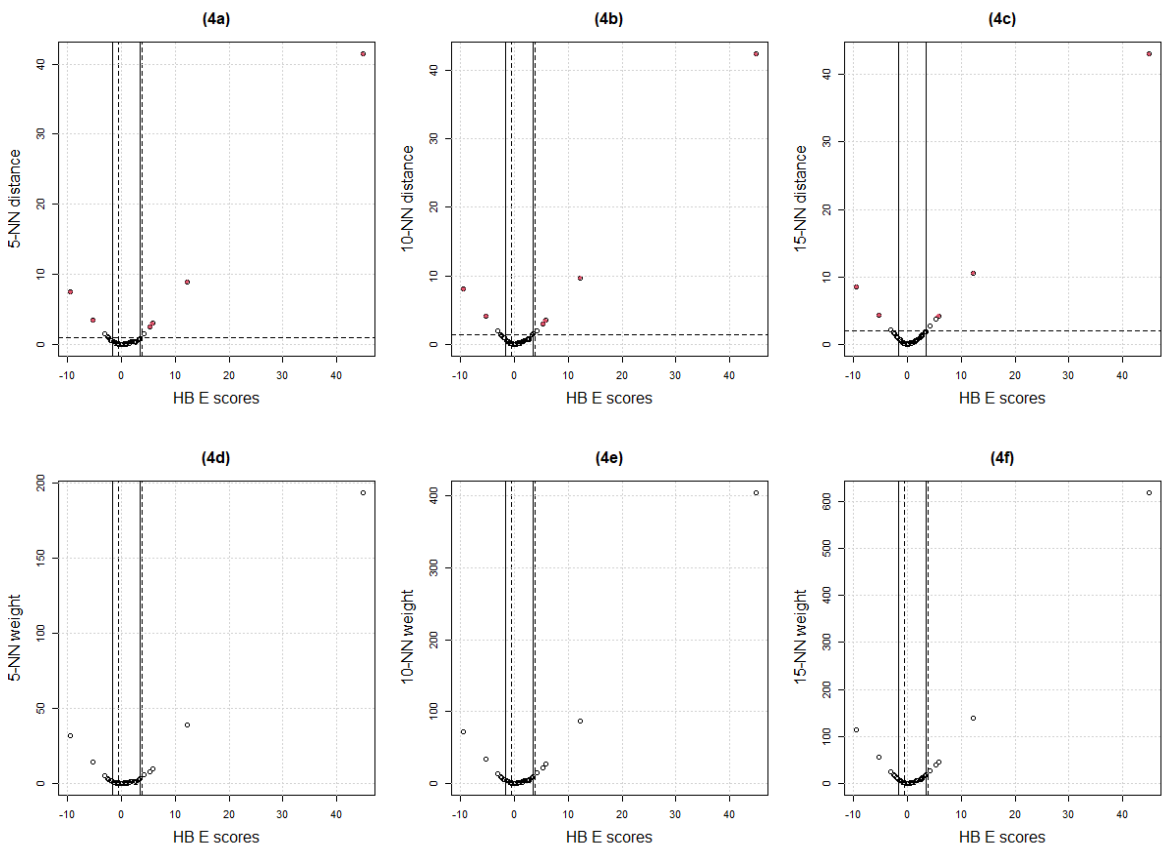

Table 4 shows that also in this case the IF score is the one with higher correlation (measured in terms of Kendall's tau) with the absolute value of the HB scores ($|E_i|$). Rankings based on IF scores tend to agree more with those provided by 10-NN and 15-NN methods. In general, correlations are all quite high.

**Table 4 – Kendall's correlation between the scores assigned to the farms producing rice**

|            | IF     | 5-NN-dist | 10-NN-dist | 15-NN-dist | 5-NN-weight | 10-NN-weight | 15-NN-weight |
|------------|--------|-----------|------------|------------|-------------|--------------|--------------|
| $|E|$      | 0.8627 | 0.7320    | 0.8285     | 0.8306     | 0.7064      | 0.8205       | 0.8401       |
| IF         |        | 0.8084    | 0.9080     | 0.9125     | 0.7798      | 0.9054       | 0.9299       |
| 5-NN-dist  |        |           | 0.8071     | 0.7934     | 0.8978      | 0.8775       | 0.8306       |
| 10-NN-dist |        |           |            | 0.9154     | 0.7624      | 0.9105       | 0.9480       |
| 15-NN-dist |        |           |            |            | 0.7595      | 0.8890       | 0.9449       |
| 5-NN-weight|        |           |            |            |             | 0.8333       | 0.7910       |
| 10-NN-weight|       |           |            |            |             |              | 0.9392       |

Figures 5 and 6 show the results obtained by applying the investigated outlier detection methods when analysing the change in individuals' wages from year 1977 to 1978, reported in Wages dataset. Plot (5a) shows that there is an increase of the wage for a large subset of individuals. The distribution of the HB $E_i$ scores is positively skewed ($M = 0.3162$) leading to identification of relatively few outliers; in this case, since there is a high concentration of the $E_i$ around the median, in expression (4) it was decided to replace $E_{Q1}$ and $E_{Q3}$ with respectively $E_{P10}$ and $E_{P90}$, as suggested by Hidiroglou and Emond (2018). This is the reason for the large discrepancy between the HB bounds and those provided by the SABP. Outliers identified by the HB method are individuals with an IF score greater of equal than 0.7. DBSCAN returns the same outliers identified by HB, with the exception of $g = 6$ ($k = 5$) (scatterplot 6b) where some additional individuals are identified as outliers. In general, scores provided by the methods based on *k*-NN show clearly identifiable potential outliers that generally correspond to those identified by the HB procedure.

**Figure 5 – Scatterplot of the individuals' wages (5a), distribution of the HB scores (5b) and relation between HB and IF scores (5c)**

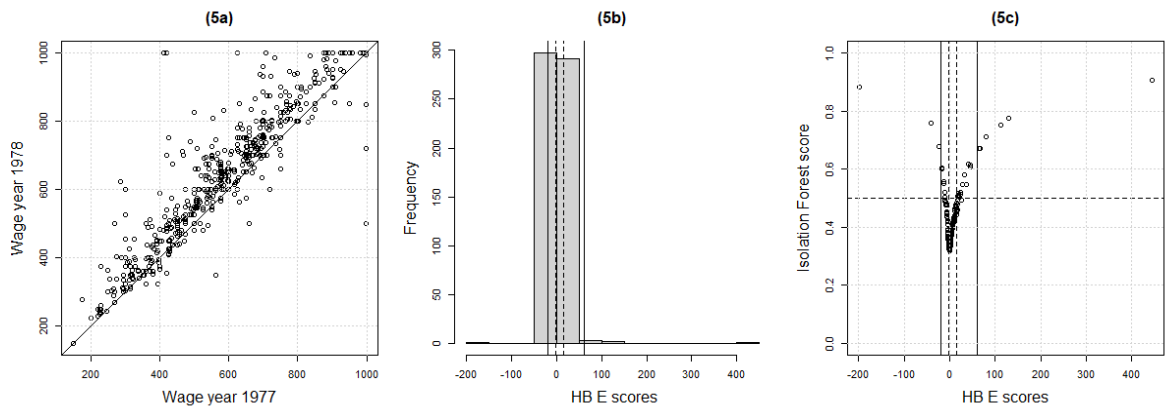

**Figure 6 – Wages data, relationship between HB and scores provided by the *k*-NN methods**

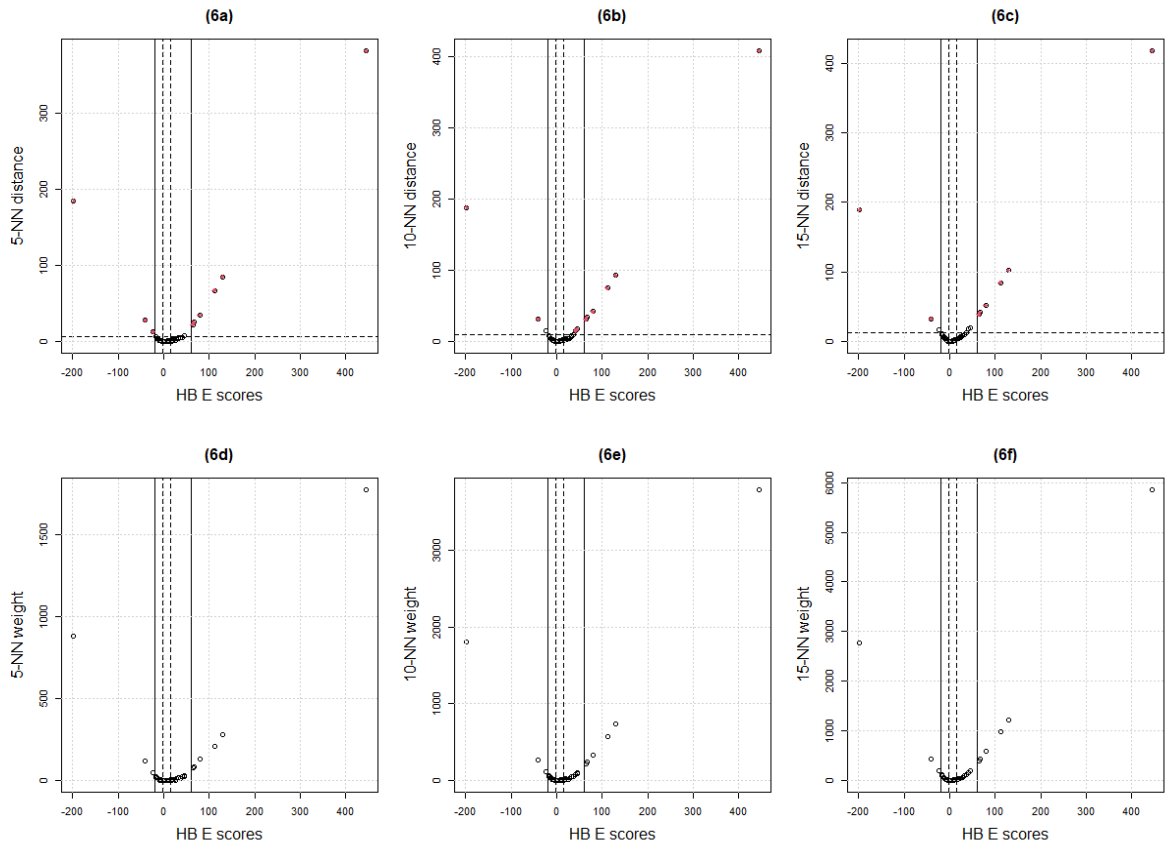

On average, the estimated correlations reported in Table 5 are lower than those estimated with other datasets, indicating that in this case the rankings provided by the scores do not fully agree. As in other cases the IF scores are those with higher correlation with the starting $|E_i|$.

**Table 5 – Kendall's correlation between the scores assigned to the individual wages**

|  | IF | 5-NN-dist | 10-NN-dist | 15-NN-dist | 5-NN-weight | 10-NN-weight | 15-NN-weight |
|---|---|---|---|---|---|---|---|
| \|E\| | 0.8101 | 0.5775 | 0.7045 | 0.7698 | 0.5420 | 0.6654 | 0.7254 |
| IF |  | 0.6376 | 0.7695 | 0.8371 | 0.6068 | 0.7426 | 0.8061 |
| 5-NN-dist |  |  | 0.7287 | 0.6678 | 0.8617 | 0.8377 | 0.7629 |
| 10-NN-dist |  |  |  | 0.8396 | 0.6712 | 0.8727 | 0.9200 |
| 15-NN-dist |  |  |  |  | 0.6225 | 0.7861 | 0.8803 |
| 5-NN-weight |  |  |  |  |  | 0.7836 | 0.7088 |
| 10-NN-weight |  |  |  |  |  |  | 0.8976 |

Figure 7 and 8 summarize the analyses done on household consumption observed in years 2014 and 2016 for the panel component of the Survey on Household Income and Wealth (SHIW) carried out by the Bank of Italy (the survey is biennial). Histogram shows $E_i$ having a symmetric distribution ($M = -0.024$) and the HB bounds ($C = 7$ and $A = 0.5$) determine a quite high number of outlying households. Similarly, the number of households with IF scores greater than 0.6 is non-negligible, but this subset reduces significantly if the threshold $u_0 = 0.7$ is set.

DCSCAN clustering procedure (plots 8a-8c; with $\delta = 30$, $\delta = 55$ and $\delta = 65$, respectively; decided by after graphical inspection of the sorted *k*-NN distances) returns a number of outliers much smaller if compared to the HB method. More in general, all the plots related to *k*-NN distances or *k*-NN weight (Fig. 8d-8f) return final scores that facilitate the identification of the outlying observations.

**Figure 7 – Scatterplot of the household consumption (7a), distribution of the HB scores (7b) and relation between HB and IF scores (7c)**

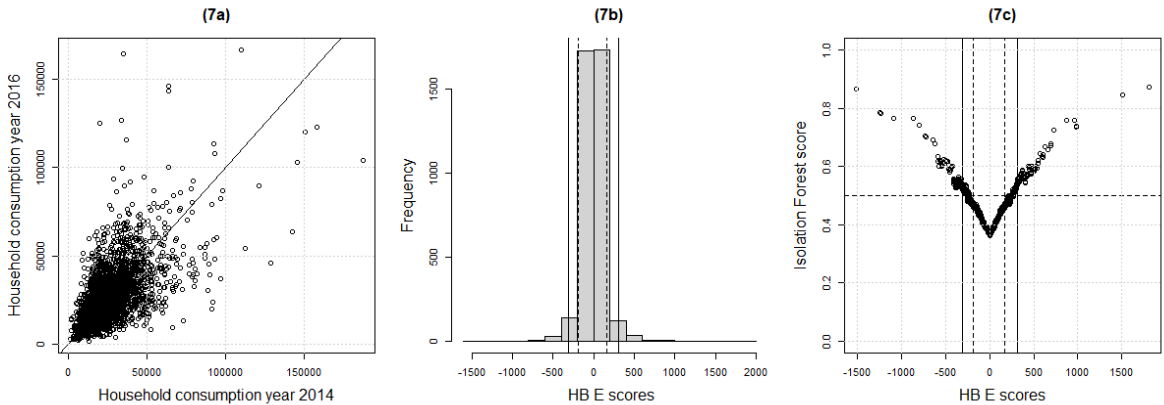

**Figure 8 – Household consumption data, relationship between HB and scores provided by the *k*-NN methods**

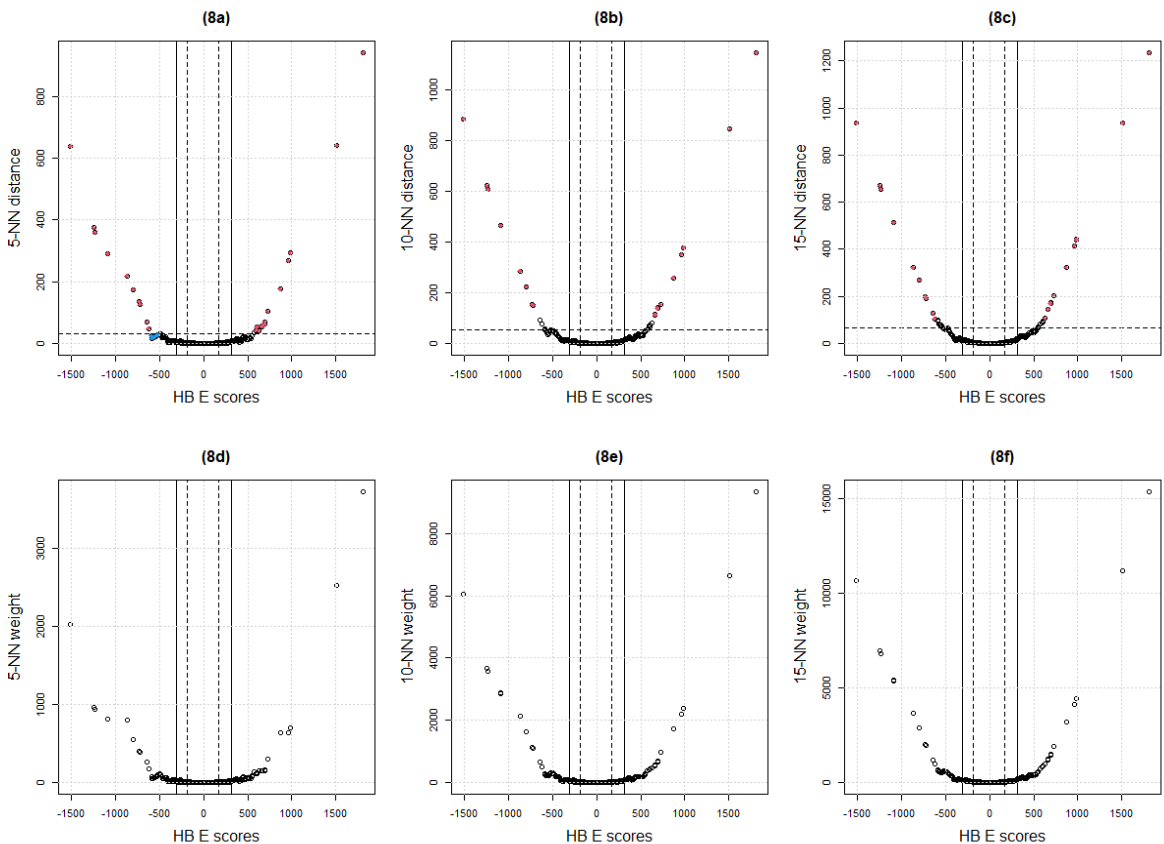

Kendall's correlations in Table 6 show the same tendency highlighted in other situations.

**Table 6 – Kendall's correlation between the scores assigned to the individual household consumption**

|  | IF | 5-NN-dist | 10-NN-dist | 15-NN-dist | 5-NN-weight | 10-NN-weight | 15-NN-weight |
|---|---|---|---|---|---|---|---|
| \|E\| | 0.9340 | 0.6702 | 0.7519 | 0.7896 | 0.6493 | 0.7268 | 0.7668 |
| IF |  | 0.7048 | 0.7877 | 0.8232 | 0.6833 | 0.7663 | 0.8083 |
| 5-NN-dist |  |  | 0.7751 | 0.7417 | 0.8737 | 0.8677 | 0.8158 |
| 10-NN-dist |  |  |  | 0.8610 | 0.7390 | 0.8820 | 0.9180 |
| 15-NN-dist |  |  |  |  | 0.7129 | 0.8204 | 0.8965 |
| 5-NN-weight |  |  |  |  |  | 0.8406 | 0.7845 |
| 10-NN-weight |  |  |  |  |  |  | 0.9169 |

It is worth noting that all scatterplots comparing the HB and IF scores show a "V" shaped diagram with the exception of Figure (1c) (firms' production) where the $E$ scores show an asymmetric distribution with moderate negative skewness ($M = -0.2338$) but quite "log" tails. The rule of thumb that identifies as potential outliers the units with an IF score greater than 0.5 in SHIW and firms' datasets returns a quite high fraction of potential outliers compared to others, this outcome seems to suggest that such a rule should be applied carefully and not in an automatic manner.

When comparing the HB $E$ scores with those provided by $k$-NN and "$k$-NN weight", the scatterplots show a kind of "U" shaped curve with some irregularities depending on the asymmetry in the distribution of the $E$ scores; an exception is again shown by the firms' production data (Fig. 2). In general, all these scatterplots exhibit some differences when passing from $k = 5$ to $k = 10$, while shapes remain almost the same for $k = 10$ and $k = 15$ (obviously the magnitude of the distance-based scores increases by increasing the values of $k$) indicating that increasing too much the value of $k$ may not be useful. DBSCAN is closely related to $k$-NN since $g = k + 1$, and the analysis of the $k$-NN distances is required to identify a threshold (parameter $\delta$); it is not a simple task and we opted for a subjective choice guided by a graphical inspection instead of using expression (8) which would require setting the additional tuning constant $\varepsilon$; it is worth noting that for each of the considered datasets the obtained results remain almost stable when varying the combination of the tuning parameters ($g$ and $\delta$); more in general, it seems that this approach returns a relatively small number of observations having however a high chance of being outliers.

## 4. Conclusions

This paper compares traditional and recent approaches to detecting outliers in longitudinal data, which is a relatively simple situation consisting of the application of univariate outlier detection methods. The traditional approaches considered in this study — the HB method and the boxplot — are popular in official statistics because they allow potential outliers to be identified directly (i.e. units outside the estimated bounds). The HB method requires the setting of a series of tuning parameters, which depend on the observed distribution of the scores ($E_i$), derived by transforming the initial ratios ($r_i = y_{t_2 i}/y_{t_1 i}$). The method assumes an approximate Gaussian distribution for $E_i$, allowing for slight skewness; however, choosing the tuning parameters (to derive the $E_i$ and the final bounds) may require several attempts. The skewness-adjusted boxplot does not explicitly assume a distribution for $E_i$ (apart from working with an unknown unimodal distribution) and allows for moderate skewness. However, it is not sufficiently flexible, as the bounds become too narrow for empirical distributions showing very long tails.

Of the wide range of nonparametric methods for outlier detection that have been developed in the fields of data mining and machine learning, we believe that those based on k-NN distances and isolation forests are efficient and capable of handling panel survey data collected by NSIs. These methods are more flexible than traditional ones since they can adapt to different empirical distributions. They assign a score to each observation, whereby the higher the score, the greater the likelihood of being an outlier. However, this is also their main disadvantage, as it is up to the practitioner to set a threshold such that units with a score beyond it are identified as potential outliers. DBSCAN provides a clear identification of outliers, but setting a threshold for the distance in addition to the value of g is required. In the case studies considered in this comparison, the chosen combination of input parameters generally returns a smaller number of potential outliers than traditional approaches and k-NN. For these reasons, DBSCAN seems preferable to k-NN methods, particularly as it allows "non-standard" distributions of $E_i$ to be captured.

In the case of an isolation forest, setting the initial tuning parameters is simpler, as the practitioner only needs to decide on the size of the bootstrap sample and the number of trees to grow, guided by the rule of thumbs mentioned in the literature. The isolation forest has the additional advantage of producing scores in the [0, 1] interval, with the midpoint (0.5) representing a suitable initial threshold value.

In general, the great advantage of "new" nonparametric methods is that they are designed to work in multidimensional settings too, unlike the boxplot and the HB. This is an appealing feature in official statistics, where data sources often include many variables collected on the same set of units. However, additional investigation is required to better understand the pros and cons of these relatively new nonparametric methods.